# Analytical Model for Light Scattering in Transparent Composites

Bin Chen[1,*], Lars A. Berglund[1], Sergei Popov[2]
Email: binchen@kth.se

1. Wallenberg Wood Science Center, Department of Fibre and Polymer Technology, KTH Royal Institute of Technology, SE-10044 Stockholm, Sweden

2. SCI school, Applied Physics Department, KTH Royal Institute of Technology, SE-11419 Stockholm, Sweden

Transparent composites that combine optical transmittance with mechanical performance are increasingly important for applications in optical devices, sustainable building materials, and photonic engineering. However, predicting light scattering in such materials remains a challenge due to complex, multi-scale microstructural interactions. Here, we present a physically grounded and computationally efficient analytical model. It predicts angular light scattering in transparent composites based on the Average Interface Number (AIN) — a single governing microstructural metric derived in this work from geometrical optics. The model accurately captures angular scattering behavior in both fiber- and particle-reinforced composites, as well as in transparent wood. We further introduce the Equivalent Average Interface Number (EAIN), combining AIN with refractive index mismatch into a unified parameter for fast haze prediction. Deep neural network (DNN) analyses confirm AIN as the dominant feature influencing optical scattering. The model predictions are supported by ray-tracing simulations and experimental trends from literature. Finally, we demonstrate the application of our model in fast image rendering simulations through transparent composites. This work provides a compact and practical toolbox for optical design and optimization of transparent structural materials.

## 1. INTRODUCTION

Transparent composites [1–7], such as wood fiber-polymer [6,7], MXene-cellulose [5], and glass fiber-resin [2,3] composites, combine optical transparency and good mechanical properties. They are typically prepared by reinforcing the transparent matrix, such as polymer, glass, ceramics, with refractive-index (RI) matched transparent fillers, such as fibers and particles. These materials are promising in optical applications where mechanical performance is important, such as glasses in electronic displays, vehicle windshields, and large-scale windows in buildings [8].

Optical performance of transparent composites is typically evaluated in terms of three interrelated metrics: transmittance (total light passing through materials), haze (light scattered at large angles in the forward direction), and opacity (light blocked, absorbed, or diffusely scattered). Accurate capturing of the angular scattering is critical for controlling haze, which in turn determines image clarity in practical applications. A primary task in preparing transparent composites is reducing haze and opacity. An important reason for haze in transparent composites is the refractive index mismatch between the matrix and the fillers.

Several transparent composites have been developed. A transparent bioinspired nacreous composite [1] prepared by filling micrometer-sized nacreous glass tablets in refractive-index-matched polymethyl methacrylate (PMMA) matrix shows superior fracture toughness, decent strength, and high optical transmittance. A nacre-like composite [8] can be prepared by connecting glass platelets via mineral bridges and infiltrating with a refractive-index matched polymer matrix, leading to high fracture toughness and good optical transparency. To obtain good optical transmittance, an organic dopant, phenanthrene, is added to PMMA to carefully refine the refractive index of PMMA, until it matches the RI of the particles [1,8].

Glass-fiber-reinforced polymer (GFRP) composites can reach a strength of 333 MPa while maintaining a good optical transmittance [9–12]. Wood fibers can also be used as reinforcement for transparent composites after delignification or chromophore removal. The holocellulose fiber network can be filled by PMMA, leading to more sustainable transparent biocomposites with a modulus of 20 GPa, a strength of 310 MPa, and transmittance of 90% at a thickness of 95 μm [13]. RI-matched polymer can also be a matrix in delignified wood, resulting in transparent wood (TW) composites [14–18]. The well-preserved wood microstructure (cellular structure) provides very good mechanical performance. The PMMA can be replaced by RI-matched poly (limonene-acrylate), leading to fully bio-based transparent wood [19]. Materials design for controlled optical scattering and transmittance is critical in the development of these transparent composites.

The complex and highly coupled microstructural parameters in these composites mean that predictions of optical scattering are challenging. The microstructural parameters include fiber diameter, sample thickness, fiber volume fraction, and fiber wall thickness for hollow fibers, etc.. The scattering can be experimentally measured by integrating sphere devices [20,21]. For modeling, numerical simulation is suitable for studies on the effect of different microstructural parameters. Kienle and Wetzel [22] numerically studied the optical properties of biological tissues containing aligned cylindrical and spherical scatterers using Monte Carlo methods. Optical properties of native wood were also studied by this method [23].

The optical scattering of transparent wood was also investigated by finite element simulation [24], where a mesh of the real 2D cross-sectional image was used. This method is not well suited for parametric studies of microstructural parameter effects on optical scattering, since structural parameters from real 2D cross-sectional images cannot be readily adjusted. As an alternative, a parametric study was performed by numerically generating different 3D wood microstructures using systematic variations in microstructural parameters, followed by ray tracing without the time-consuming meshing of a 3D model [25]. The study revealed that the optical scattering of transparent wood is affected by multiple coupled parameters, such as sample thickness, fiber size, fiber cell wall thickness, existence of vessels, fiber volume fraction, etc.. The coupling of these parameters makes it challenging to predict the optical scattering directly from these structural parameters.

Analytical models can quantitatively predict the optical properties of transparent composites. An empirical equation [26] shows that the light transmittance of transparent glass particle-reinforced polymer (GPRP) composite is a function of the interface surface area and the refractive index difference in glass-polymer. This empirical equation has two unknown empirically defined parameters lacking direct physical interpretation. A quantitative method [27] was proposed to determine the transmittance of fibre-reinforced transparent composites. However, it is developed for the composites with constant fiber size. It cannot predict the scattering angle distribution, which is critical in imaging or display optics. Neither can they evaluate the optical haze (transmitted light scattered at a large angle) of a composite.

We have not been able to find a quantitative analytical model in the literature to predict the scattering angle distribution of transmitted light in transparent composites. A reliable model should offer simplicity but needs to be based on the physics of the problem, and its predictive ability needs to be verified. Here, we propose an analytical model for scattering angle distribution in transparent composites where the light beam is orthogonal to unidirectional cylindrical fibers (Figure 1). It is derived based on geometrical optics [28] (Figure 1 (a)) and predicts angle distribution directly from microstructural images. The model reveals that the average interface number (AIN) along the light propagating direction is the main structural parameter governing optical scattering in composites.

We validated this model for composites with unidirectional fibers or with spherical particles (Figure 1(b, c)). This model was also used to study the optical scattering of TW composites, since the fibers are mostly unidirectionally aligned along the longitudinal direction. A set of 3D microstructural models of TW was generated (Figure 1(d)). Optical scattering was then numerically simulated for 10,000 2D cross-section images of TW, where the optical scattering is very significant. The analytical model predictions correlate well with the numerical simulation results for TW composites (Figure 1(e)). A single metric, the equivalent average interface number (EAIN) combines the average interface number AIN with the RI difference between polymer and reinforcement. This EAIN number was proposed for simple and efficient haze prediction in TW (Figure 1(e)) by a single parameter. A deep neural network (DNN) [29] was trained to predict the haze in transparent wood, and the training error is used to validate the critical role of AIN on optical scattering (Figure 1(f)). Finally, the proposed analytical model demonstrated its potential in a computer vision application for image rendering simulations (Figure 1(g)).

**Analytical model development and validation**

(a) Analytical model development  (b) Light scattering simulation  (c) Analytical model validation

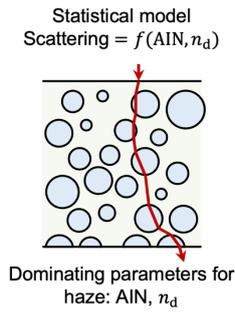
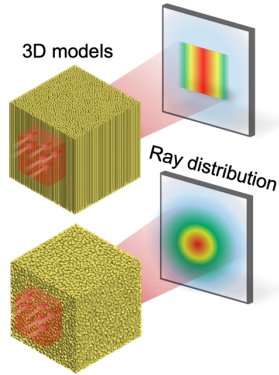
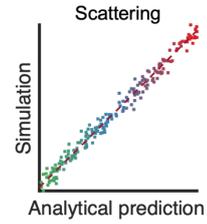

**Scattering study for light scattering of transparent wood composites**

(d) Dataset generation  (e) Analytical model for TW

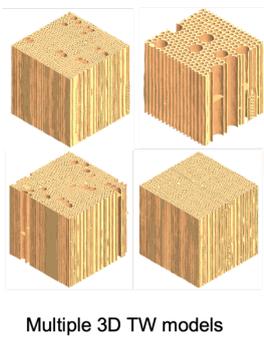
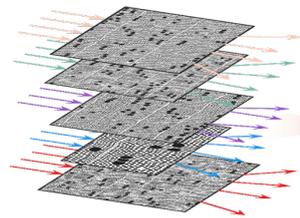
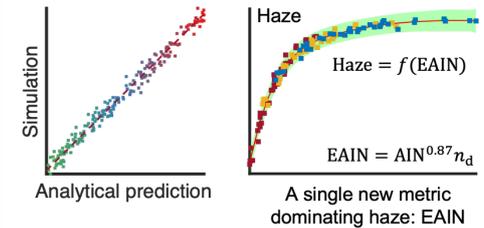

(f) Deep learning: Haze prediction & parameter evaluation

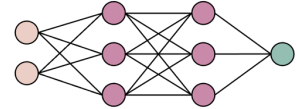

**Application**

(g) Virtual camera simulation using analytical model

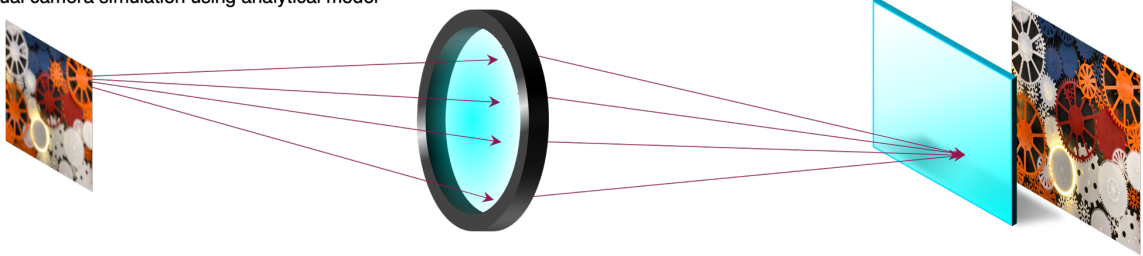

Figure 1. Schematic diagram of this work.

## 2. RESULTS

### 2.1 Analytical model

A key finding in this study is an analytical expression (Eq. (1)) describing the scattered angle distribution of an optical beam traveling through transparent composites. The scattering is presented as a function of a structural parameter called average interface number (AIN), as well as the refractive index difference between reinforcement and matrix. The assumptions for the analytical model include:

- The reinforcements, such as fibers and particles, are randomly distributed position-wise in the cross-sectional plane.
- The composites are produced without defects, and the refractive index mismatch between matrix and reinforcement is the only cause of the optical scattering.
- The reinforcements have random sizes at the order of micrometers, and thus, geometric optics can be applied.

For clarity, the derivation of Eq. (1) is detailed in further sections, but here we emphasize the key steps:
a) Modeling ray deflection at a single fiber under the geometrical optics assumption.
b) Computing angular variance through multiple fibers.

c) Linking this variance to structural parameters via AIN.

In geometrical optics, the ray deflection (Figure 2(a)) occurs only at the interface between the matrix (RI is $n_m$) and the reinforcement (RI is $n_r$). The standard deviation $\sigma(\Theta_e)$ of the emergent ray angle $\Theta_e$ through a transparent composite is analytically obtained (see detailed derivation in Section 4.4) as follows:

$$\sigma(\Theta_e) \approx \sqrt{-\frac{1}{2}\ln\left(1 - n_r^2(1 - e^{-M\mathrm{Var}(\Delta\theta_i)})\right)} \quad (1)$$

where $M$ is the AIN along the ray propagating direction, which can be calculated from the cross-section images. $\theta_i$ is the incident angle (Figure 2(b)) of the single fiber or particle. $\mathrm{Var}(\Delta\theta_i)$ is the variance of $\Delta\theta_i$ (the angle direction change after passing through a fiber or particle). The analytical model can be used for different composites, where $\mathrm{Var}(\Delta\theta_i)$ varies with the refractive index difference $n_d = n_m - n_r$ and the fiber or particle geometry, which should vary accordingly for different reinforcements.

The analytical model shows that the most critical structural parameters dominating the light scattering through transparent composites are AIN ($M$). Other parameters, including sample thickness, fiber size, volume fraction, etc., have an indirect impact on light scattering. The scattering can be predicted by resorting to only one instead of a large set of structural parameters. This analytical model agrees well with the empirical equation in literature [26], where the particle surface area, a parameter highly correlated with the AIN for 2D models, is considered as the key structural parameter.

In this work, we take the composites with cylindrical reinforcement as examples, such as the GFRP and TW composites. The cylindrical fibers as the reinforcements are assumed to be unidirectionally aligned but randomly distributed position-wise in the cross-sectional plane. The determination of the variance of $\mathrm{Var}(\Delta\theta_i)$ of a single cylindrical fiber is presented in Section 4.5. Based on Eq. (1), $\sigma(\Theta_e)$ versus AIN and $|n_d|$ is shown in Figure 2(c) and Figure 2(d), respectively. It increases almost linearly with $|n_d|$ at a constant AIN $M$=50 (Figure 2(d)), while nonlinearly with AIN (Figure 2(c)).

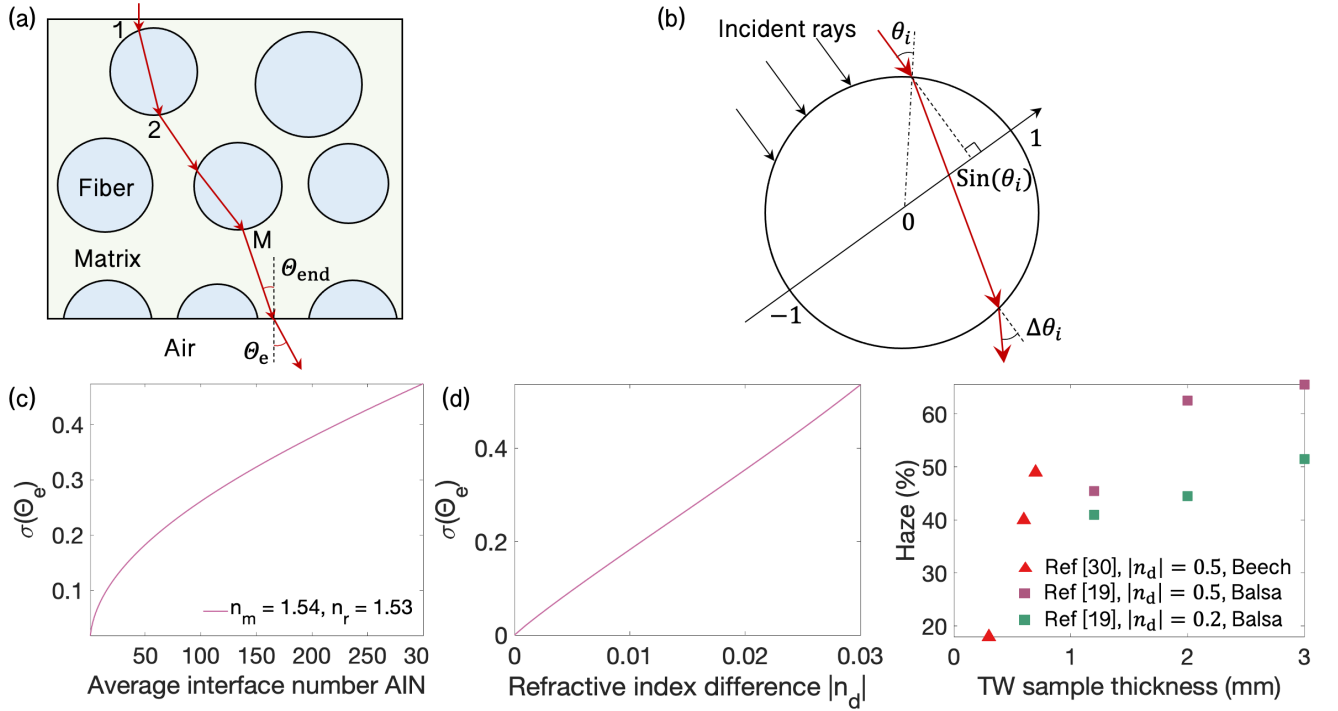

Figure 2. Analytical scattering model: (a) principal diagram of ray scattering in the cross section of a transparent composite; (b) ray deflection of a single fiber; analytical result of $\sigma(\Theta_e)$ versus (c) AIN and (d) refractive index difference $|n_d|$ when $n_m = 1.54$ and $M = 50$; (e) Experimental haze data of TW in literature [19,30] for model validation, where the wood species and refractive index difference are labelled on the figure.

Experimental validation can hardly be implemented quantitatively, since the angle distribution is difficult to measure. However, we can qualitatively validate the model by comparing the haze, which is one of the most commonly used scattering indicators, for TW composites. The haze is measured as the ratio of rays that have an angle greater than 2.5° [31]. Haze is highly correlated with the standard deviation of the emergent angle $\sigma(\Theta_e)$. A larger haze usually denotes a larger $\sigma(\Theta_e)$. The experimental haze value cannot be compared across literature due to the differences of processing procedures, material sources, and measurement methods, but we can compare the haze in the same literature. The experimental results in [19,30] both show that the scattering, haze, increases with sample thickness (also AIN, since the wood species is the same in each literature). The experimental results in Ref. [19] demonstrate that the scattering, haze, also increase with the refractive index difference $|n_d|$.

## 2.2 Model validation and refinement using glass-fiber-reinforced polymer composites

The analytical model was applied to unidirectional GFRP (Figure 3(a)). A dataset including 1,000 2D microstructures with different thickness and structural parameters was generated numerically (Section 4.1). The two material phases in the composite have a small RI mismatch. The ray paths in the composites were traced (Section 4.3) and the emergent ray angles were obtained. A statistic of the total reflection (TR) event number (Figure S6) reveals that there is a significant number of TR events when $n_d < 0$. For the same $|n_d|$ and AIN, the simulation shows similar TR event numbers (Figure S6) when $n_d < 0$ and $n_d > 0$. It demonstrates the effectiveness of the simplification of the single fiber scattering from Eq. (8) to Eq. (9).

For each microstructural image, $\sigma(\Theta_e)$ was obtained (Figure 3(b)) by using both numerical simulation (Section 4.3) and the proposed analytical model. Each point in the figure corresponds to one material with a different microstructure. The slope of the fitted line in Figure 3(b) is 0.937, which is close to the ideal line (the dashed line with slope of 1), demonstrating a good agreement between analytical predictions and numerical simulation results. The overestimation in the analytical model occurs mainly at larger scattering $\sigma(\Theta_e)$. It demonstrates that the model works better when the scattering $\sigma(\Theta_e)$ is approximately below 0.5.

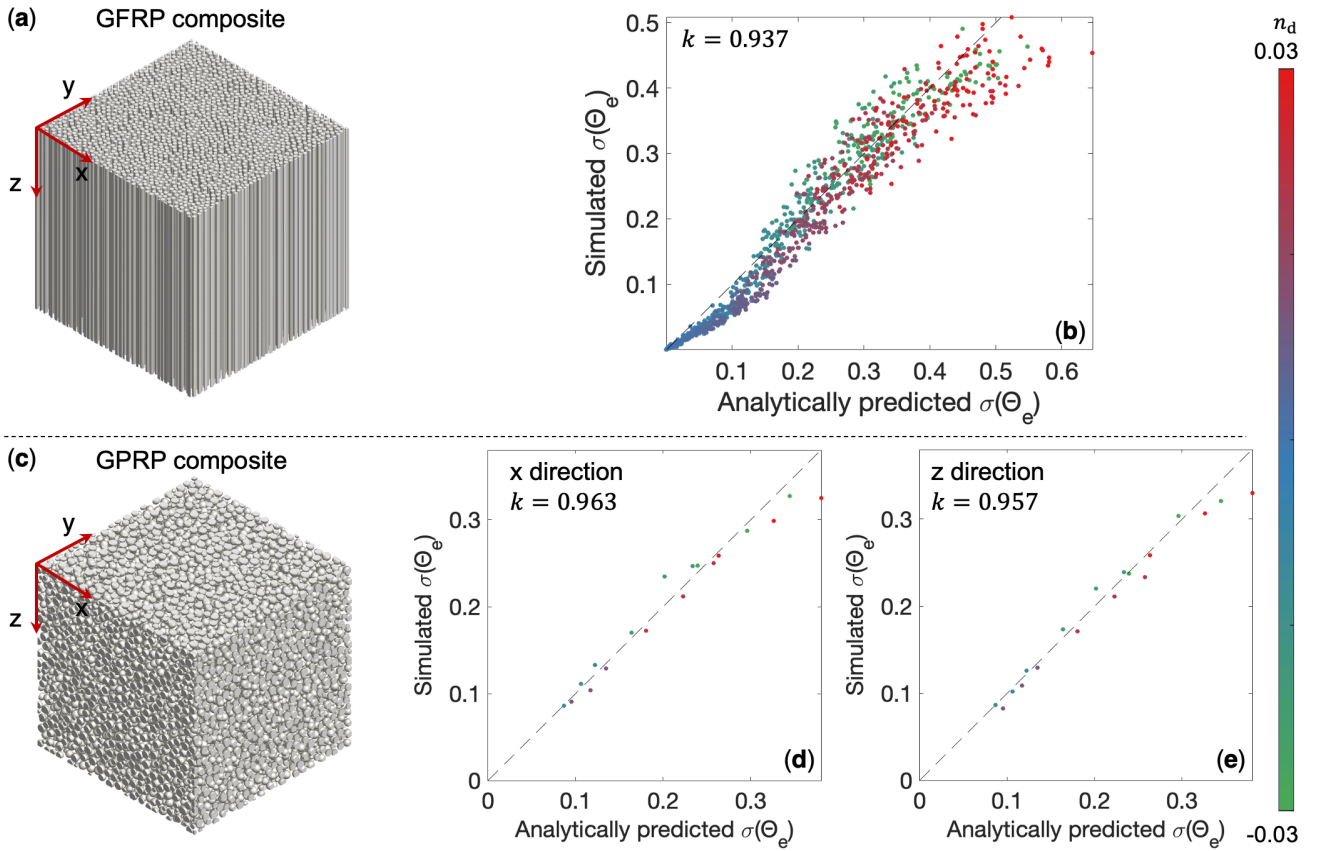

Figure 3. Analytical model validation on simulated (a, b) GFRP (fibers) and (c-e) GPRP (particles) composites: (a) A 3D microstructure model of GFRP composites; (b) $\sigma(\Theta_e)$ in simulation versus analytical prediction using Eq. (1) and Eq. (9); (c) a 3D microstructure model of GPRP composites; comparison of the $\sigma(\Theta_e)$ obtained by simulation and analytical prediction in (d) x direction and (e) z direction for GPRP composites.

## 2.3 Model application on simulated glass-particle-reinforced polymer composites

The developed model was also validated by predicting the scattering in GPRP composites (Figure 3(c)). The ray paths in the composites were traced using ray tracing (Section 4.3). The projection angle of the emergent ray in x and z directions (Figure 3(c)) can be obtained. Then, $\sigma(\Theta_e)$ in both x (Figure 3(d)) and z (Figure 3(e)) directions were obtained. The model cannot perfectly predict the scattering, since the analytical model was developed on a 2D instead of a 3D microstructure model. Deviations can arise due to the 3D scattering paths where curvature-induced total internal reflection is not fully captured by the 2D formulation. The projections of a critical angle in the x and z directions are smaller than the critical angle, and thus may be without total reflection. However, the errors are in an acceptable range. The slopes $k$ of the fitted line of the data points (Figure 3(d) and Figure 3(e)) are around 0.96, which is close to 1, indicating a good match between simulation and model prediction.

## 2.4 Light scattering in transparent wood composites

### 2.4.1 Ray scattering prediction using the analytical model

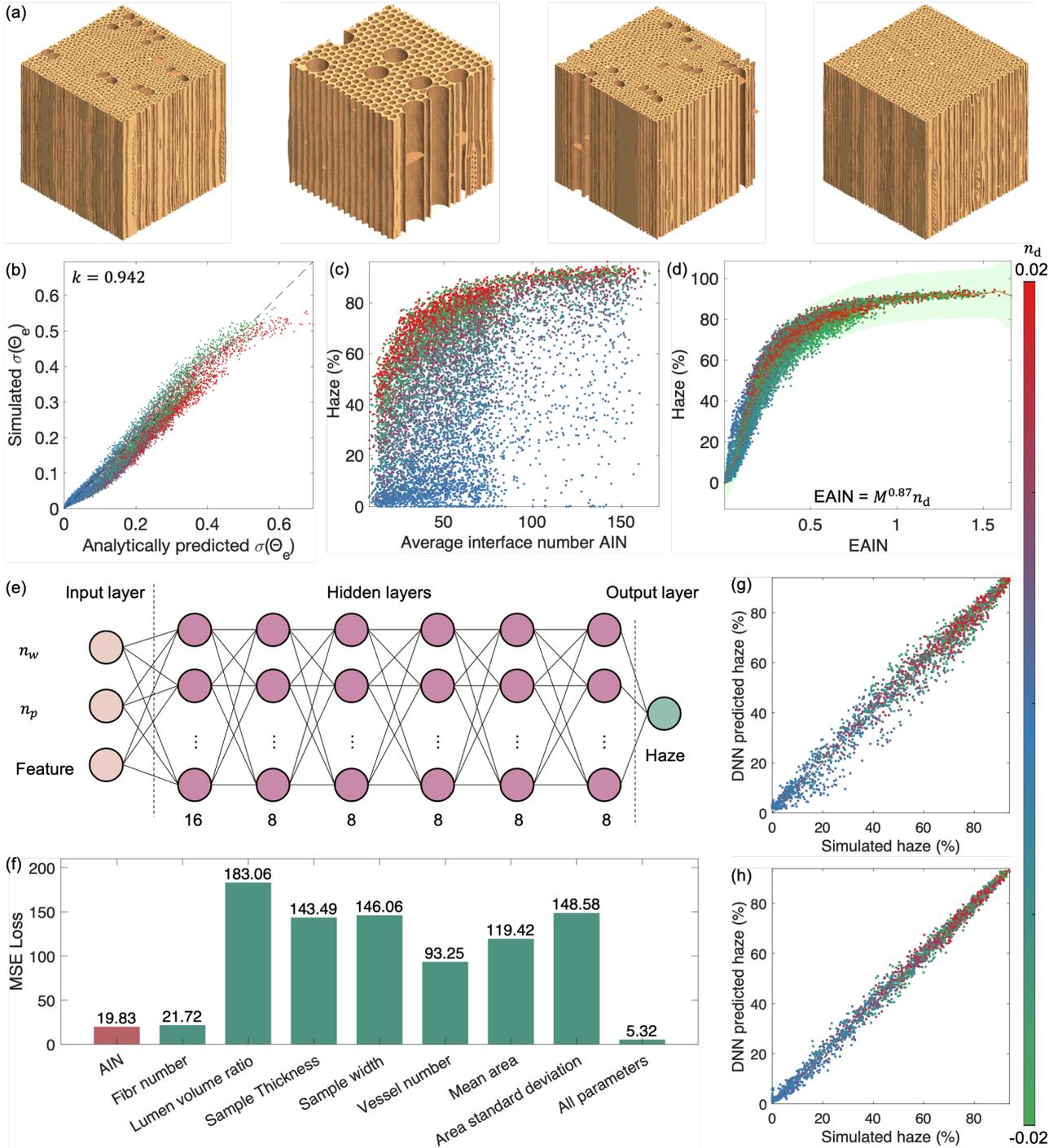

Figure 4. Optical scattering study in TW composite. (a) examples 3D TW microstructures numerically generated by different structural parameters; (b) theoretically predicted light scattering $\sigma(\Theta_e)$ versus the simulated ones; (c) Haze with respect to AIN and $n_d$; (d) Haze versus $EAIN = M^{0.87} n_d$ fitted by a 10-th order polynomial function, where the green band is the residual band; (e) DNN architecture for haze prediction with different features; the input layer includes one given feature plus the RIs of the two compositions. The DNN with all features as input layer is also constructed; (f) the MSE loss on test set for DNN with different input features; DNN predicted haze versus the simulation-derived haze on the test set for (g) the network with only AIN and RIs as inputs and (h) the network with all parameters as inputs. The color indicates the value of $n_d$.

We also validated the proposed analytical model on the simulated TW composites (Section 4.2) [32]. We extract 10,000 2D microstructural models from multiple 3D microstructure models (see four examples with the size of $1000 \times 1000 \times 1000$ voxels in Figure 4(a)). The simulated scattering $\sigma(\Theta_e)$ (scattering along radial direction) is in good agreement with the analytically predicted data (Figure 4(b)). This provides confidence in the approximate analytical model. The fitting of the data gives a slope value of 0.942, which is close to the ideal value 1. The slight difference indicates an overestimate of the scattering by the analytical model for TW. It might be a result of the existence of ray cells in TW. The ray cells increase the AIN, but do not obviously contribute to the scattering along the radial direction. The results again show that the analytical model overestimates the scattering when $\sigma(\Theta_e) > 0.5$.

2.4.2  Equivalent Interface number for haze prediction in transparent wood

We also studied the haze of TW. Simulation demonstrates that the haze increases with respect to $|n_d|$ (Figure S1) and AIN (Figure 4(c)). The analytical model shows that two parameters AIN and $n_d$ govern the scattering, considering an approximation constant $n_r$. Here we introduce a new single indicator, called equivalent average interface number (EAIN) - empirically constructed as a function of AIN and RI mismatch, that governs the haze with a single parameter. It is defined as $\text{EAIN} = M^k |n_d|$, where $k \in [0,1]$ is an unknown coefficient. One benefit is that it can further reduce the number of governing parameters from 2 to 1.

To find the optimal coefficient $k$ for EAIN, we fit the haze as a function of EAIN using a polynomial function with orders varying from 5 to 25. For each polynomial function order, we can find the optimal $k$ and polynomial function coefficients that best fit the data points. The optimal values of $k$ for different polynomial function orders are quite stable (Figure S5(a)), while the root mean square errors of the haze in the fitting are shown in Figure S5(b). The optimal $k$ is obtained as $0.87 \pm 0.01$. Therefore, the EAIN can be obtained as

$$\text{EAIN} = M^{0.87} |n_d| \qquad (2)$$

The haze-EAIN curve fitted by a 10$^{\text{th}}$-order polynomial function is shown in Figure 4(d). Video S1 shows the effect of $k$ on the fitting and residual error. All data points are located in the vicinity of the curve, with the green area the residual band for polynomial fitting. It demonstrates that the haze is highly dependent on this new single parameter EAIN, or in other words, EAIN is a governing parameter for haze. We can compare the haze by comparing only EAIN.

2.4.3  Haze prediction in transparent wood using deep neural networks

The analytical model reveals that AIN and RIs of the two phases in TW are the key parameters for the haze. However, the relationship between these parameters and haze is still unknown. The impact of other parameters on the haze has not been evaluated. The DNN (Figure 4(e) for the architecture) is used herein to predict haze by using some typical TW parameters, including cell wall RI, polymer RI, AIN, lumen volume ratio, average fiber cross-section area, standard deviation of the fiber cross-section area, sample thickness (in pixels) and width, vessel and fiber number. Different combinations of the input features are studied in the network to evaluate their impact on the haze. The RIs of the two counterparts are always part of the input features. The third feature is chosen from the remaining set, sequentially, to construct a three-feature input layer (Figure 4(e)). Additionally, a DNN with all features as inputs is also trained as a benchmark. Consequently, several different DNNs with the same architecture, but different inputs, are constructed. The loss of the haze prediction on the test set is selected as a metric to show the importance of each feature on the haze of TW.

Figure. S4(a, b) shows the convergence history of the DNN with AIN and the two RIs as input layers, and all features as input layers, respectively. The AIN-based DNN (three input features, the other two are the RIs) gives much smaller (79% to 89%) training loss than the DNNs based on the other input features (Figure 4(f)), except for the one with fiber number and the one with all parameters. The small mean square error (MSE) loss of the DNN with AIN and RIs as input features confirms that AIN is the most critical microstructural parameter for the haze. The DNN with fiber number as inputs also leads to a small training loss. However, the fiber number is not a critical parameter for the haze since the images may have varied widths. For instance, keeping the thickness but changing the width of the TW sample may significantly change the fiber number, but does not affect the scattering at all.

The trained DNN can directly predict the haze using several structural parameters as inputs. The DNN predictions match well with the simulated results (Figure 4(g)), if only RIs and AIN are used as inputs. The DNN can better predict the haze (Figure 4(h)) if all features are included in the input layer due to the enriched input information. Considering the simplified model (3 input features versus 10 input features), the error in Figure 4(g) is still reasonable.

It further demonstrates that AIN is a governing parameter for scattering in TW.

## 2.5 Analytical model for image rendering simulation through a transparent composite

In the art design and product development using transparent composites, it is very helpful to visualize the blurring effects of the composites in advance. We can simulate that very efficiently (Section 4.7) using backward ray tracing and the proposed analytical model (Figure 5(a)). Images of a sample behind a transparent composite can be numerically rendered by a virtual camera or eye. The object with a size of 20×20 mm (left of Figure 5(b)) was positioned behind the transparent GFRP composite at distances $d$ of 10 and 20 mm.

Instead of tracing the ray paths in the complex 3D microstructures, which is very time-consuming, we can sample the emergent ray orientation following the distribution given by the analytical model. The computational burden is negligible compared with the ray tracing in 3D transparent composite microstructures. It can also avoid using a large microstructure model, which is very time-consuming to generate, for a large piece of TW sample. The rendered images are shown in Figure 5(b). The larger distance between the transparent composite and the object leads to more significant image blurring. The increased AIN also yields more significantly blurred images. Note that here we considered only scattering along the horizontal direction using the analytical model, while the scattering along the fiber orientation was neglected.

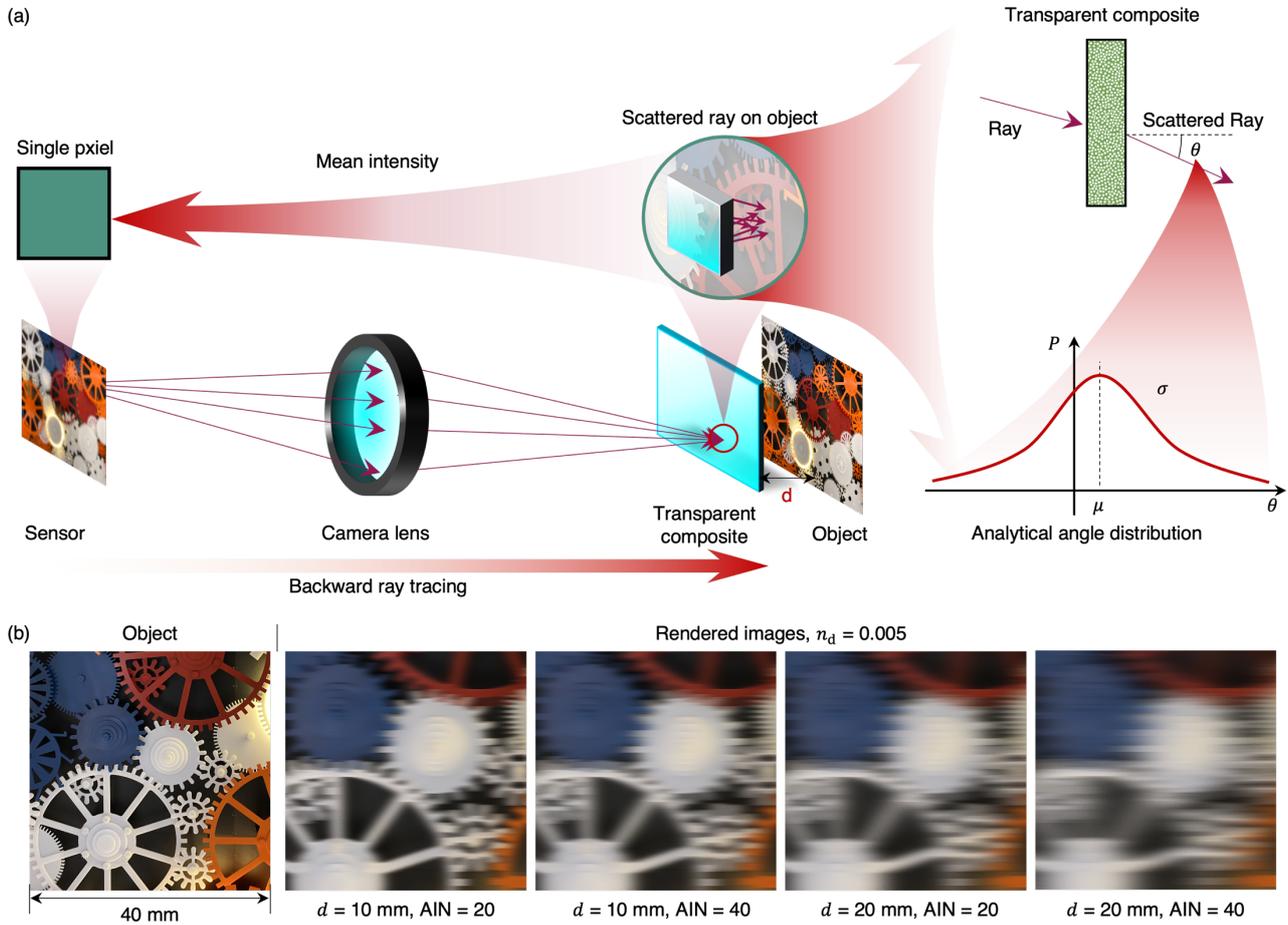

Figure 5. Image rendering simulation through GFRP composite using ray tracing and the analytical model: (a) principle and (b) the blurred images rendered by a virtual camera with the distance between transparent composite and the object of 10 mm, 20 mm, and AIN of 20, 40, respectively.

## 3. DISCUSSION AND CONCLUSION

This work develops an analytical model for quantitatively predicting the ray angular scattering resulting from refractive index mismatch in transparent fiber-polymer composites. The model is developed based on the following assumptions:

1)  Fibers have a cylindrical shape with diameters of micrometers, such that the geometrical optics can be

applied;
2) Cylindrical fibers are aligned longitudinally and randomly distributed in the cross-section plane.
3) Materials have a homogeneous structure without obvious defects, such as air bubbles inside.

We validated the analytical model on both GFRP and TW composites with numerical simulation, which demonstrates a good match. Though this model is not developed for random 3D material models, it still shows good angular scattering prediction accuracy for 3D GPRP composites.

The model reveals that the average interface number (AIN) along the thickness direction and the refractive index difference are the key parameters governing the optical scattering. The other structural parameters, such as fiber size, fiber diameter, volume fraction, sample size, sample thickness, etc., are not explicitly included into consideration to contribute to the optical scattering. Instead, these parameters affect the optical scattering by changing the AIN. This finding is further confirmed by the DNN model. The DNN models with different structural parameters and the refractive indices as inputs were trained to predict the haze. The DNN models with only three parameters (average interface number and refractive indices of the two material phases) can well predict the haze with a small training error, while those with the other structural parameters will lead to larger training errors. While we did not conduct experiments in this study, we validated the model against experimental data reported in the literature, and the trends are consistently reproduced. Future work will aim to extend the model with dedicated experimental validation.

The new parameter, AIN, further inspired the development of a new metric, denoted as the equivalent average interface number (EAIN). This new parameter combines the impact of AIN and refractive index difference for a rough estimation of haze from a single parameter. This compact and efficient method considerably simplifies new material design in an intuitive way. For instance, one can develop transparent wood composites with smaller scattering by selecting wood species and polymer with a smaller EAIN.

This work also demonstrates the potential of this analytical model in image rendering simulation through a transparent GFRP composite. It can significantly speed up the image rendering in numerical simulation by avoiding the ray tracing in complex 3D micro-structured models. The simulation shows that the larger the distance between the transparent composite and the object, the more significant the extent of image blurring. Future work may explore extending the analytical model to account for anisotropic RI distributions, irregular reinforcement shapes.

## 4. METHODS

### 4.1 Dataset for microstructure models of GFRP and GPRP

The GFRP and GPRP composites were numerically generated by modifying an open-source script [33], which was originally designed for drawing randomly distributed and non-overlapped circles of various sizes. The original script works only for 2D cases. The 3D GFRP composite model can be obtained by extruding the 2D model along the thickness direction. The 3D GPRP composite model was simulated by modifying the script from 2D to 3D. Note that the physical size of the microstructure in this work is not important, which is in units of pixels.

The generation of a large number of microstructures is time-consuming. We generated 20 cross-sections of the GFRP microstructure with a dimension of 3000×3000 pixels. A set of smaller images was then clipped from the large images at random locations and with random sizes to form a dataset with 1000 samples. The clipped images have thickness (ray propagating direction) varying from 1000 to 3000 pixels and a constant width (perpendicular to the incident rays) of 3000 pixels. The AIN along the thickness direction for those images is calculated between 15.4 to 74.9. These smaller images can be considered as the cross-sections of different GFRP microstructures. The RI for the fibers was randomly selected between 1.48 and 1.54, while the RI difference was selected between -0.03 and 0.03.

One large 3D GPRP composite with a size of 2000×2000×2000 voxels is also generated. Sphere particles of different sizes are randomly distributed. A set of smaller microstructures with different sample thicknesses (500, 1500, and 2000 voxels) was then clipped from the large 3D model, which is a more efficient way to generate a dataset. The AIN varies from 13.7 to 27.3. The RI of the matrix was fixed as 1.54, while the RI difference between the matrix and particles was selected from -0.03 to 0.03 with an increment of 0.01. In total, 18 3D composite structures were generated.

**4.2 Dataset generation for TW models**

A large dataset of the TW model is obtained using the parameters in Table S1. The distortion-map based method proposed previously [32] is used to generate high-fidelity 3D microstructure models. However, it is very time-consuming to obtain a large number of 3D microstructures. This work obtains 10 different wood microstructures with the size of $3000 \times 3000 \times 2000$ voxels using different structural parameters, such as different fiber sizes, cell wall thickness, the existence of vessels, and the existence of ray cells. Considering that the scattering is much more significant along the radial direction than along the longitudinal direction, scattering data on a 2D model is used for simplicity. From the ten 3D wood microstructures, we sampled 10,000 cross-sectional images with different thicknesses and locations. The sample thickness was randomly selected between 501 to 2991 pixels. The images were also randomly flipped either horizontally or vertically to augment the dataset size. The RI of polymer is randomly chosen between 1.48 and 1.54, while $n_\mathrm{d}$ is set in [-0.02, 0.02].

**4.3 Ray tracing for optical scattering**

The scattered ray angle distributions were obtained using numerical simulation. In this work, the fiber and particle were assumed to have a size of micrometers. Therefore, geometrical optics can be used for numerical simulation. The RI difference between the reinforcement and matrix was assumed to be the only factor affecting the optical scattering.

Ray tracing, a powerful method in the computer graphics community [34] is adopted to simulate how the rays travel inside these transparent composites (Figure S2(a) for 2D model and Figure S2(b) for 3D model ). The ray paths inside the 2D and 3D microstructure models of the composites were traced [25]. The method used in this work is efficient since it can avoid the time-consuming structure meshing step, which is especially important for the ray tracing in 3D model. The ray can propagate forward along the same orientation if the former point and current point (with a distance of 2 pixels or voxels) are located in the same optical medium. Otherwise, it will change the orientation. In the same way, we can track the location of the ray after every two pixels along the optical path until the ray exits the microstructure. Then, we can obtain the complex light path inside the microstructure.

A beam consisting of 1000 rays is incident perpendicularly on one surface of the simulated composite. The rays are randomly distributed in the middle region with a size of 500 pixels for 2D models (GFRP and TW, Figure S2(a)), and a region with a size of 500×500 pixels for 3D models (GPRP, Figure S2(b)). Note that in TW, the rays are incident in the radial-longitudinal plane and travel along the tangential direction. In each case, the beam can cover multiple fibers or particles, enabling statistically meaningful results in the ray tracing simulation. After undergoing scattering through the complex microstructure of the composites, the rays emerge from the opposite side. The rays that are emitted from the side surfaces were neglected. Once we know the standard deviation of the angular scattering $\sigma(\Theta_\mathrm{e})$ of all the emergent rays on the opposite surface, we can use it to indicate the optical scattering of a sample. A detailed method for ray tracing can be found in previous work [25].

**4.4 Scattering of transparent composites**

Suppose the reinforcements are aligned cylindrical fibers. The rays are assumed to meet, on average, $M$ the interface number (or M/2 fiber) in the microstructure (Figure 2(a)) along ray propagating direction. Suppose the incident angle to $i$th fiber is $\theta_i$; the angle change of the ray at $j$th fiber is $\Delta\theta_{i,j}$. The angle $\Theta_\mathrm{end}$ of a ray after passing through all fibers can be written as

$$\Theta_{\text{end}} = \sum_{j=1}^{\frac{M}{2}} \Delta\theta_{i,j} \tag{3}$$

Suppose that the fibers are randomly distributed; then the incident angle of the rays at each fiber is independent and randomly distributed. The angle change $\Delta\theta_{i,j}$ of different fibers should also be independent. Then, the variance of $\Theta_{\text{end}}$ (Figure 2(a)) can be written as

$$\text{Var}(\Theta_{\text{end}}) = \frac{M}{2}\text{Var}(\Delta\theta_i) \tag{4}$$

where $\text{Var}(\Delta\theta_i)$ is the variance of the angle deflection through a single fiber, which is derived in Supporting Information S.II. Note that $\Theta_{\text{end}}$ is the angle corresponding to the emergent light of the last fiber. The rays, still, need to leave TW sample to air, where the RI between air and the composite is significant. It is reasonable to assume that $n_m$ is equal to $n_r$, when we study the refraction from TW to air environment. The RI of air is supposed to be 1. The variance of $\sin(\Theta_e)$ can be roughly estimated as

$$\text{Var}(\sin(\Theta_e)) = n_r^2 \text{Var}(\sin(\Theta_{\text{end}})) \tag{5}$$

Based on the variance of the sine function (Supporting Information S.IV), we have

$$\frac{1}{2}\left(1 - e^{-2\text{Var}(\Theta_e)}\right) \approx \frac{n_r^2}{2}\left(1 - e^{-2\text{Var}(\Theta_{\text{end}})}\right) \approx \frac{n_r^2}{2}\left(1 - e^{-M\text{Var}(\Delta\theta_i)}\right) \tag{6}$$

The scattering can then be derived as

$$\text{Var}(\Theta_e) \approx -\frac{1}{2}\ln\left(1 - n_r^2\left(1 - e^{-M\text{Var}(\Delta\theta_i)}\right)\right) \tag{7}$$

The standard deviation $\sigma$ of $\Theta_e$ then has the form of Eq. (1). The $\sigma(\Theta_e)$ can be obtained once we have the variance of the angle deflection through a single fiber $\text{Var}(\Delta\theta_i)$ (Supporting Information S.II and S.VII).

### 4.5 Optical scattering through a single cylindrical fiber

For the transparent composites with cylindrical fiber reinforcement, $\text{Var}(\Delta\theta_i)$ of a single cylindrical fiber is derived in Supporting Information S.II. It has the form of

$$\begin{cases} \text{Var}(\Delta\theta_i) = 4\left(1 - \frac{|n_d|}{n_m}\right)\text{Var}\left(\theta_i - \arcsin\left(\frac{n_m}{n_r}\sin(\theta_i)\right)\right) + 4\frac{n_d^2}{n_m^2}, & n_m > n_r \\ \text{Var}(\Delta\theta_i) = 4\text{Var}\left(\theta_i - \arcsin\left(\frac{n_m}{n_r}\sin(\theta_i)\right)\right), & n_m < n_r \end{cases} \tag{8}$$

where $n_d = n_m - n_r$ is the RI difference between the matrix and reinforcement. The difference of $\text{Var}(\Delta\theta_i)$ between the cases of $n_m < n_r$ and $n_m > n_r$ is from the existence of total reflection when $n_m > n_r$, which may lead to a large $\Delta\theta_i$. We assume the rays are uniformly distributed in the orientation vertical to the ray propagation direction (Figure 2(b)) for each fiber. It denotes that $\sin(\theta_i)$ is uniformly distributed in [-1, 1] when $n_m < n_r$, and in $\left[\frac{|n_d|}{n_m} - 1, 1 - \frac{|n_d|}{n_m}\right]$ when $n_m > n_r$. The variance $\text{Var}\left(\theta_i - \arcsin\left(\frac{n_m}{n_r}\sin(\theta_i)\right)\right)$ can be numerically calculated for any given RIs considering circular cross-section (Supporting Information S.VII). Based on Eq. (8), the standard deviation of the angle deflection $\sigma(\Delta\theta_i)$ after passing through a single fiber is shown in Figure S3 when both $n_d < 0$ and $n_d > 0$.

For ideal cylinder fibers, there should not be any total reflection (TR) in the proposed analytical model when $n_d < 0$ (Figure S7). However, the imperfection of the circular fiber cross-section and error in ray tracing may lead to total reflection in the simulation, where the TR can significantly deflect the ray propagating orientation. We can simplify the assumption for $n_d < 0$ by assuming the same single fiber scattering as that of $n_d > 0$, i.e., the rays refracting in $\left[\frac{|n_d|}{n_m} - 1, 1 - \frac{|n_d|}{n_m}\right]$ and reflecting in the remaining regions for each fiber along the diameter. Based on the new assumption, the angle variation for a cylindrical fiber in Eq. (8) is refined as

$$\text{Var}(\theta_i) = 4\left(1 - \frac{|n_d|}{n_m}\right)\text{Var}\left(\theta_i - \arcsin\left(\frac{n_m}{n_r}\sin(\theta_i)\right)\right) + 4\frac{n_d^2}{n_m^2} \tag{9}$$

This Eq. (9) has a simpler form compared to Eq. (8) and will be used in this work, since it does not calculate

Var($\Delta\theta_i$) under two conditions.

**4.6 Neural network for haze prediction in transparent wood**

A DNN model (Figure 4(e)) with 6 hidden layers is constructed to predict the haze from some basic parameters. Deep learning is a powerful technique that has recently been widely used in material science [35]. In this work, we use a regression model implemented on PyTorch [36]. The dataset used in this model includes 10, 000 instances (Table S1), and each instance has several features of the TW sample. The features include $n_m$, $n_r$, AIN, lumen volume ratio, average fiber cross-section area, standard deviation of the fiber cross-section area, sample thickness (in pixels) and vessel number. The label of each instance is the haze. 80% of the instances are split as a training set, while the remaining are the test set.

This model has 7 fully connected layers with ReLU activation function [37], followed by a sigmoid activation function to constrain the haze prediction between 0 and 100. The Adam optimizer [38] is used with a learning rate set to 0.01, while the learning rate is reduced by a factor of 0.2 for every 30 epochs. In the 6 hidden layers, 16, 8, 8, 8, 8 and 8 neurons are included, respectively. The loss function used for the training task is mean square error (MSE). The training task terminates after 100 epochs (see the training error in Figure S4).

**4.7 Virtual camera simulation through GFRP**

The analytical model for optical scattering angle distribution can also help to simulate the image blurring through a transparent composite. We have done this task for transparent wood using backward ray tracing, which, however, is a computationally extensive task. The computation mainly comes from two parts: large microstructure generation and backward ray tracing inside the complex microstructure. In previous work [25], only images with a size of $51\times 51$ pixels and the object with a size of millimeters were simulated due to the heavy computational burden.

The analytical model, combined with the backward ray tracing, can simplify and speed up this task significantly. A virtual camera with a sensor resolution of 500×500 pixels and pitch size of 1 $\mu m$ is simulated. The lens of the virtual camera has a focal length of 50 mm and an aperture diameter of 2 mm. The object was put at a distance of 2550 mm from the lens. A GRRP composite sample with a thickness of 1 mm was put in front of the object (Figure 5(a)) with a gap $d$ of 10 and 20 mm, respectively. The AIN is given as 20 and 40, respectively. The RI is 1.55 and 1.54 for matrix and fibers, respectively. For every single pixel on the sensor, we trace the paths of 1000 rays from that pixel to the object. The rays cannot converge to the same point on the object due to the scattering of the transparent composite, leading to image blurring. The key is to obtain the ray orientation change through a transparent composite.

Eq. (1) suggests that we can obtain the distribution of the emergent ray angle in the cross-section plane using the analytical model we proposed. The 2D analytical model cannot be perfectly used for 3D composite. However, the rays are almost vertical to the fibers with the incident angle larger than 89.5°, and thus, the scattering along fiber orientation is very small. Therefore, we directly use the 2D analytical model by neglecting the scattering along fiber orientation. The expectation of the scattered ray angle $\sin(\theta_e)$ can be directly obtained (Supporting Information S.V) without considering the scattering, while the standard deviation $\sigma(\sin(\theta_e))$ are obtained from the analytical model (Supporting Information S.IV). For each ray we simulated, we can sample the emergent angles following the analytical model. Therefore, the rendering of each pixel on the camera sensor uses 1000 emergent rays in total. The object color at the locations where the 1000 emergent rays meet the object can be obtained using ray tracing (Figure 5(a)). The color of the pixel is calculated as the average of the intensities at those locations. In the same way, we can obtain the color of all pixels in the image. The simulation is very efficient since we do not need to trace the ray paths inside 3D transparent composite microstructures.

**DATA AVAILABILITY**

The data that support the findings of this study are available from the corresponding author upon reasonable request.

**ACKNOWLEDGEMENTS**

This work is supported by Knut and Alice Wallenberg foundation (KAW 2021.0311), Formas (2022-01231). We also greatly acknowledge the HORIZON AI-TRANSPWOOD (AI-Driven Multiscale Methodology to Develop Transparent Wood as Sustainable Functional Material) project, Grant No. 101138191, co-funded by the European Union.

**CONFLICT OF INTERESTS**

The authors declare no competing interests

# Analytical Model for Light Scattering in Transparent Composites

## (Supporting information)

Bin Chen[1,*], Lars A. Berglund[1], Sergei Popov[2]
Email: binchen@kth.se

1. Wallenberg Wood Science Center, Department of Fibre and Polymer Technology, KTH Royal Institute of Technology, SE-10044 Stockholm, Sweden

2. SCI school, Applied Physics Department, KTH Royal Institute of Technology, SE-11419 Stockholm, Sweden

## S.I. Supporting figures

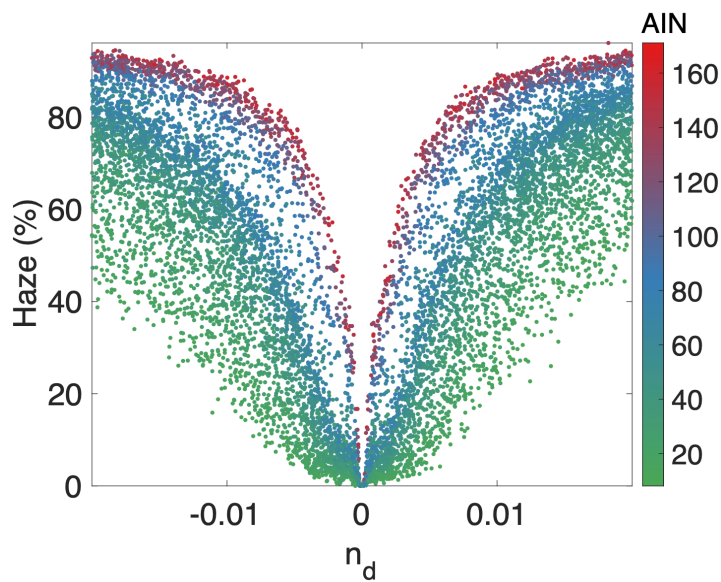

**Figure S1.** Haze versus $n_d$ in the TW. The color denotes AIN.

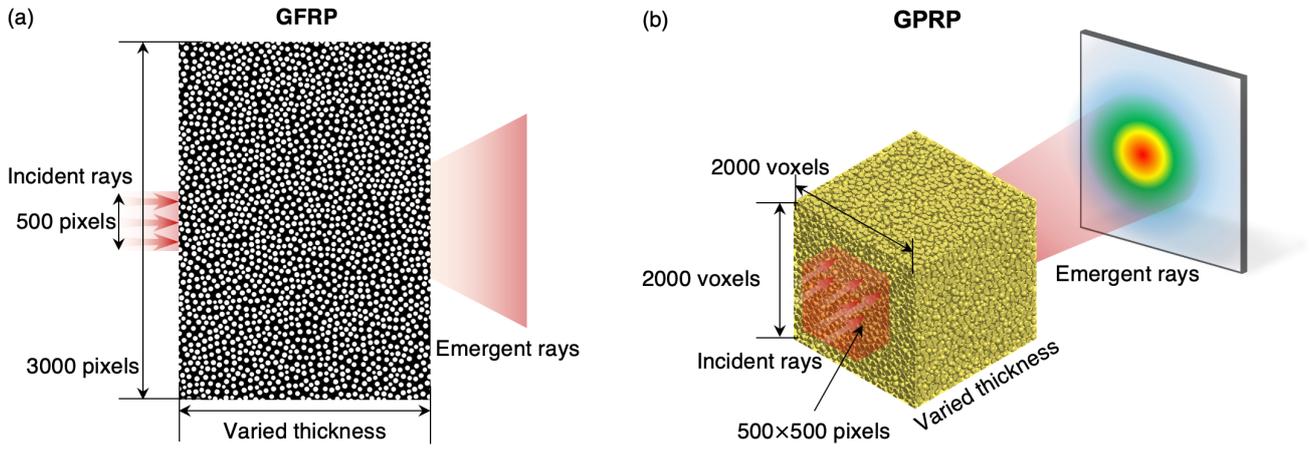

**Figure S2.** Schematic diagram for ray tracing in (a) glass-fiber-reinforced polymer and (b) glass-particle-reinforced polymer composites.

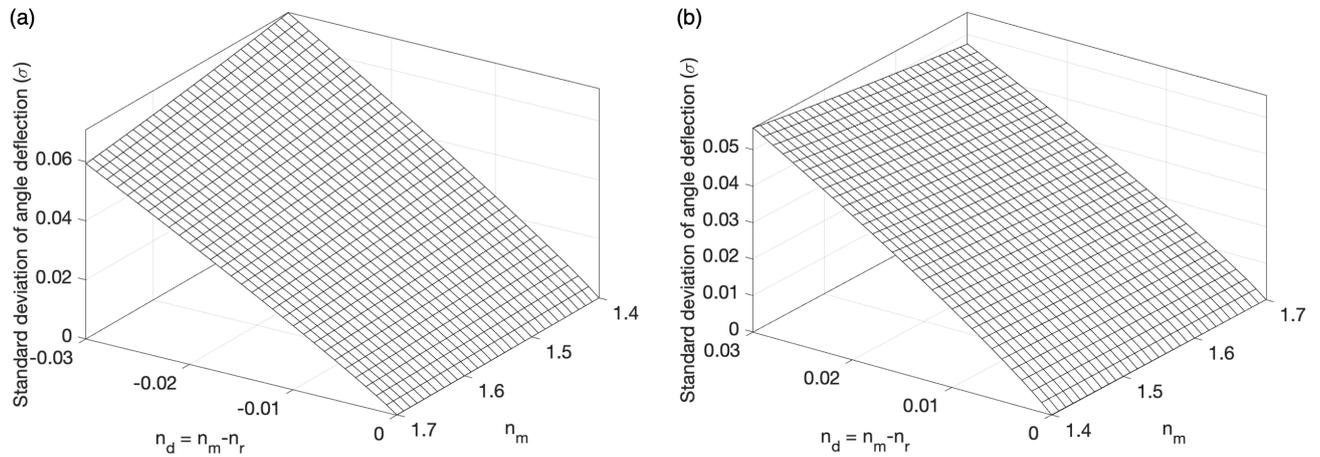

**Figure S3.** Standard deviation of angle deflection for a single fiber under the condition of (a) $n_m < n_r$ and (b) $n_m > n_r$.

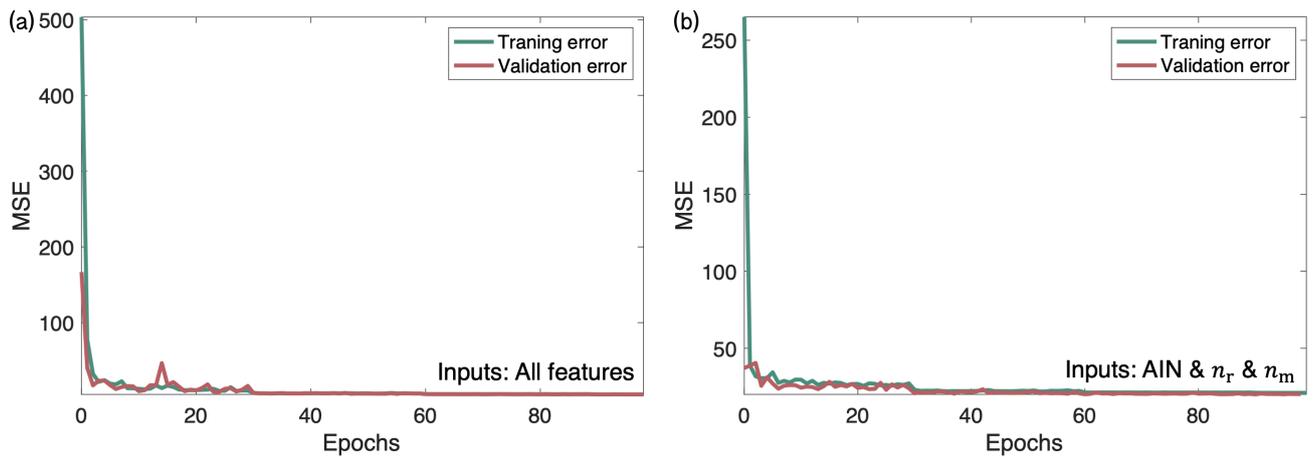

**Figure S4.** The training loss and validation loss during DNN training. (a) The input layer includes all features as input, including $n_r$, $n_m$, AIN, lumen volume ratio, average fiber cross-section area, standard deviation of the fiber cross-section area, sample thickness (in pixels), sample width, vessel number, and fiber number; (b) the input layer has only three input features, i.e., $n_r$, $n_m$ and AIN.

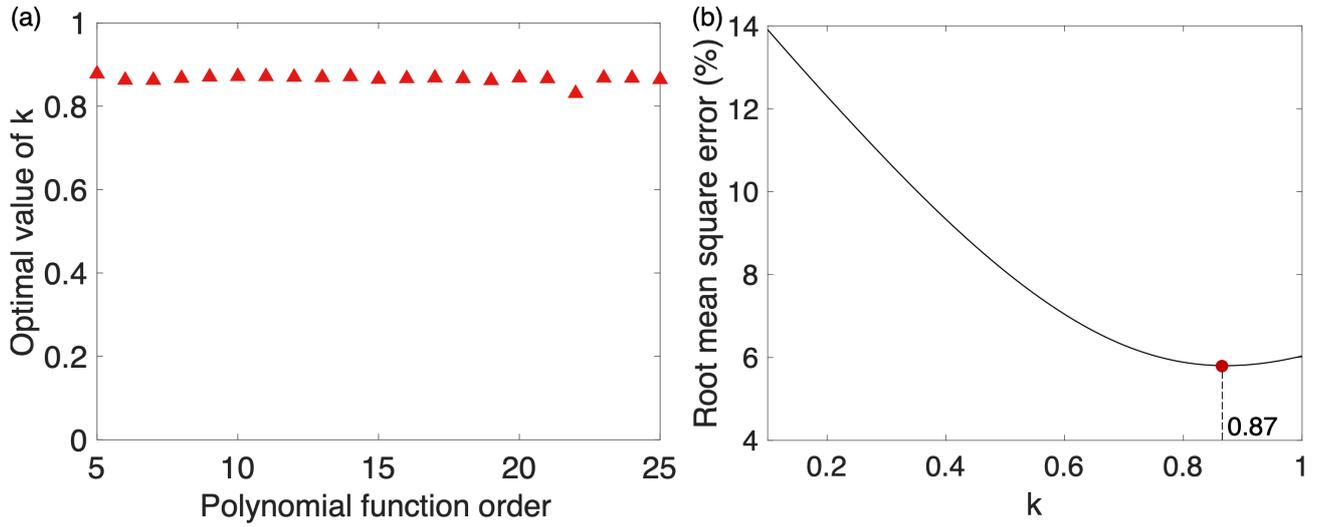

**Figure S5.** (a) The root mean square error when fitting the haze versus EAIN= $M^k|n_d|$ using $10^{th}$ order polynomial function and different $k$; (b) The optimal $k$ obtained for EAIN= $M^k|n_d|$ when different order is selected for the polynomial function.

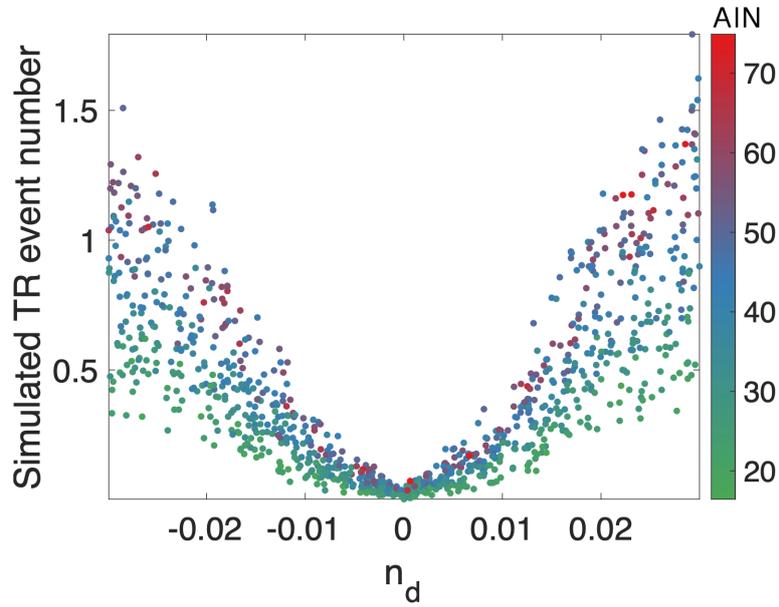

**Figure S6.** The total reflection (TR) event number at different refractive index differences $n_d$. In principle, the TR event number should be 0 when $n_d < 0$. However, the simulation shows similar values for the same $|n_d|$ and AIN.

Table S1. Parameters for the dataset generation of the 2D ray tracing model for deep learning

| Parameters | Description |
| --- | --- |
| 3D wood structure number | 10 |
| Image number | 1000 images are randomly selected per 3D wood structure models |
| Image transformation | 50% possibility to flip along each direction |
| Image size along incidental ray | 501 to 2991 |
| Image width | 1.2 times of sample thickness but smaller than the dimension of the volume size |
| Image position | Randomly located on the volume |
| $n_r$ | 1.48 to 1.54 |
| $n_d$ | -0.02 to 0.02 |
| Incidental ray number | 1000 |
| Incidental ray position | Randomly distributed in a 500×500 pixels region along radial direction. |

## S.II. Scattering of a single cylindrical fiber

The refractive index (RI) difference between the matrix (RI of $n_m$) and the fiber (RI of $n_r$) is small. For simplicity, we can reasonably neglect the reflection part at the interface (see Supporting information S.VI) if the incident angle is smaller than the critical angle. Therefore, we can track a single ray path in the composites. At the optical interface between matrix and reinforcement, there are two typical cases, $n_r > n_m$, or $n_r < n_m$. We will discuss the two cases separately.

**Condition of $n_m < n_r$**

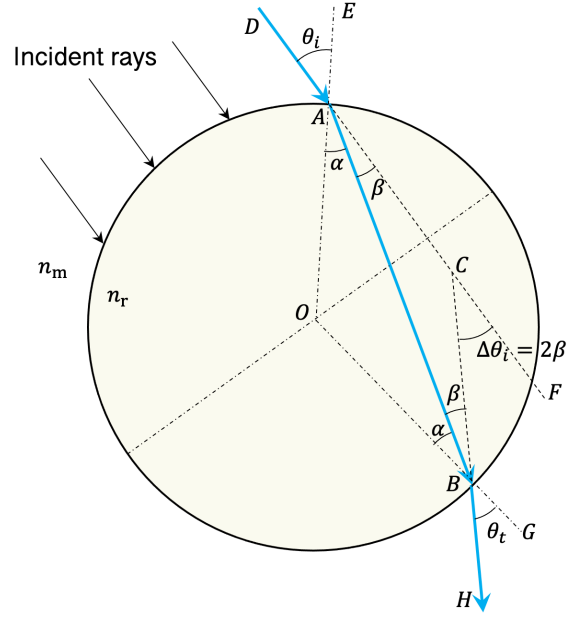

**Figure S7.** Scattering through a single circular fiber when $n_m < n_r$. The blue arrow denotes the ray. Suppose that the incident ray meets a single fiber at point A, and the incident angle is $\theta_i$. The center of the circular fiber is $O$. The ray refracts at point A. The ray then propagates inside the fiber to point B. Notate the refracted angle at A as $\alpha$. Since $|OA| = |OB|$, we have $\angle OBA = \angle OAB = \alpha$. Therefore, the refracted angle of the emergent light is $\theta_t = \angle HBG = \angle DAE = \theta_i$. Then, it yields $\angle ABC = \angle BAC = \beta = \theta_i - \alpha$. The ray angle change after passing through a single fiber should be $\Delta\theta_i = \angle BCF = \angle ABC + \angle BAC = 2\beta$.

If $n_m < n_r$, only refraction exists at each interface. As shown in Figure S7, suppose the incident angle and emergent angle at the first interface of a fiber is $\theta_i$ and $\alpha$, respectively. According to Snell's equation, it yields

$$\beta = \theta_i - \alpha = \theta_i - \arcsin\left(\frac{n_m}{n_r}\sin(\theta_i)\right) \tag{S1}$$

where $\beta$ is the angle change at the first interface. We can also obtain that the angle deflection (Figure S7) through one fiber is

$$\Delta\theta_i = 2\beta \tag{S2}$$

The variance of $\Delta\theta_i$ is calculated as

$$\text{Var}(\Delta\theta_i) = 4\text{Var}(\beta) = 4\text{Var}\left(\theta_i - \arcsin\left(\frac{n_m}{n_r}\sin(\theta_i)\right)\right) \tag{S3}$$

There is no analytical solution for $\text{Var}(\Delta\theta_i)$, which, however, can be obtained numerically (Supporting Information S.VII) for any given RIs by assuming a uniform distribution of $\sin(\theta_i) \sim U[-1, 1]$. Then we can get the estimation of $\text{Var}(\theta_{end})$.

**Condition of $n_m > n_r$**

If $n_m > n_r$, total reflection may also exist for ray deflection. In that case, we should also consider the scattering resulting from both total reflection and refraction at the interface. When passing through a single fiber, the ray may either be reflected in region 1 and region 3, or refracted in region 2 (Figure S8).

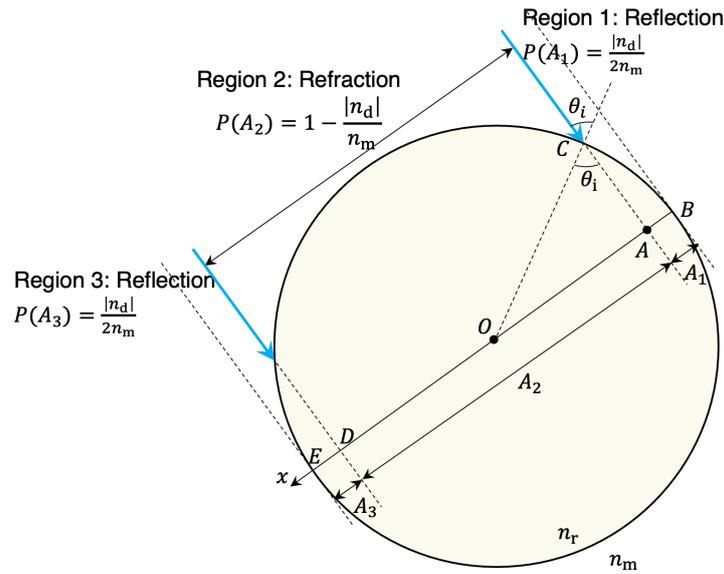

**Figure S8.** Scattering through a single fiber when $n_r < n_m$. The blue arrow denotes the ray. The fiber will be refracted in region 2, and has total reflection in regions 1 and 3. Suppose that the fiber has a radius of 1, while its center is $O$. The calculation of the possibility of the three events in the three regions is equivalent to calculating the region size, since we assume the ray is uniformly distributed in space. Suppose that the incident angle of the ray at point C is equal to the critical angle. Then we have $\sin(\theta_i) = 1 - \frac{|n_d|}{n_m}$, and $|OA| = |OD| = \sin(\theta_i)$. Thus, we also have $|AB| = |DE| = 1 - \sin(\theta_i) = \frac{|n_d|}{n_m}$. The possibility of the incident ray to meet region 1 to region 3 is $P(A_1) = \frac{|n_d|}{2n_m}$, $P(A_2) = 1 - \frac{|n_d|}{n_m}$ and $P(A_3) = \frac{|n_d|}{2n_m}$, respectively.

When the incident angle is larger than the critical angle $\arcsin \frac{n_r}{n_m}$, the total reflection occurs (Figure S8). Otherwise, it will be refraction (Figure S8).

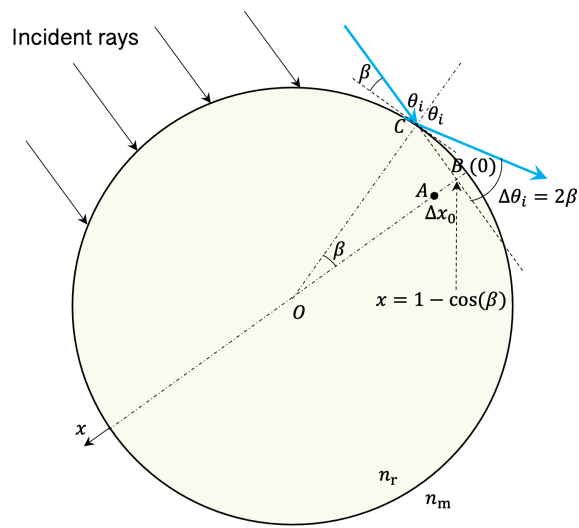

**Figure S9.** Ray deflection resulted from total reflection when $n_r < n_m$. The blue arrow denotes the ray. The incidental angle and the reflected angle should be the same for total reflection. The angle deflection after passing

through the fiber should be $\Delta\theta_i = 2\beta$. As discussed in Figure S8, the size of the total reflection region is $|AB| = \frac{|n_d|}{n_m}$. Suppose an axis with its origin at $B$ and oriented toward $O$. The coordinate of $A$ is $x_0 = \frac{|n_d|}{n_m}$.

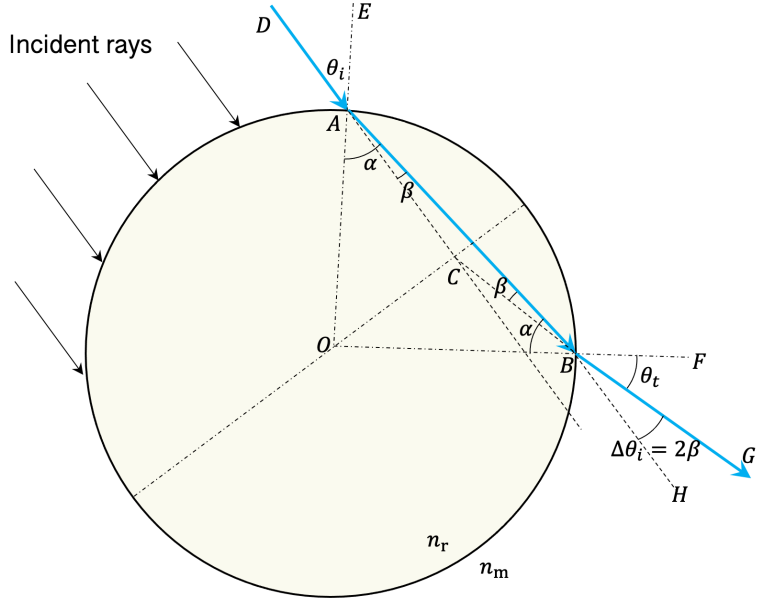

**Figure S10.** Ray deflection resulted from refraction when $n_r < n_m$. The blue arrow denotes the ray. Suppose that the incident ray meets a single fiber at point A, and the incident angle is $\theta_i$. The center of the circular fiber is $O$. The ray refracts at point A. The ray then propagates inside the fiber to point B. Notate the refracted angle at A as $\alpha$. Since $|OA| = |OB|$, we have $\angle OBA = \angle OAB = \alpha$. Therefore, the refracted angle of the emergent light is $\theta_t = \angle FBG = \angle DAE = \theta_i$. Then, it yields $\angle ABC = \angle BAC = \beta = \theta_i - \alpha$. The ray angle change after passing through a single fiber should be $\Delta\theta_i = 2\beta$.

Assume the ray passing through the three regions as three mutually exclusive and exhaustive events $A_1$, $A_2$, $A_3$ in the sample space. Since the ray is uniformly distributed along $x$ axis, the possibility for the rays in regions 1, 2, 3 is, respectively, $P(A_1) = \frac{n_d}{2n_m}$, $P(A_2) = 1 - \frac{n_d}{n_m}$, and $P(A_3) = \frac{n_d}{2n_m}$ (Figure S8). Notate the angle deflection at a single fiber as $\Delta\theta_i$.

According to the variance decomposition (Supporting information S.III), the variance of $\Delta\theta_i$ is calculated as

$$\mathrm{Var}(\Delta\theta_i) = \sum_{k=1}^{3} \mathrm{Var}(\Delta\theta_i|A_k)P(A_k) + \sum_{i=1}^{3} E(\Delta\theta_i|A_k)^2 P(A_k) - \left(\sum_{i=1}^{3} E(\Delta\theta_i|A_k)P(A_k)\right)^2 \quad (S4)$$

First, we consider the angle deflection for total reflection in region 1. For any incident angle, the angle deflection is $\Delta\theta_i = 2\beta$ (Figure S9). Region 1 has a size of

$$x_0 = \frac{n_d}{n_m} \quad (S5)$$

In region 1, select a local coordinate system with $x$ axis vertical to the incident rays (Figure S9) and the origin at the boundary of the fiber. The $x$ coordinate of the incident ray is assumed to be uniformly distributed. It has

$$x = 1 - \cos(\beta) \approx 1 - \left(1 - \frac{\beta^2}{2}\right) = \frac{\beta^2}{2} \quad (S6)$$

Then we have

$$\beta = \sqrt{2x} \quad (S7)$$

The variance of $\beta$ is obtained as

$$\mathrm{Var}(\beta) = \mathrm{Var}(\sqrt{2x}) = E(2x) - E(\sqrt{2x})^2 \quad (S8)$$

Since $x$ is uniformly distributed in $[0, x_0]$, $E(2x)$ can be calculated as
$$E(2x) = x_0 \tag{S9}$$

The expectation $E(\sqrt{2x})$ is estimated as
$$E(\sqrt{2x}) = \frac{2\sqrt{2}}{3}\sqrt{x_0} = \frac{2\sqrt{2}}{3}\sqrt{\frac{n_d}{n_m}} \tag{S10}$$

The expectation $E(2\beta)$ can be calculated as
$$E(2\beta) = 2E(\sqrt{2x}) = \frac{4\sqrt{2}}{3}\sqrt{\frac{n_d}{n_m}} \tag{S11}$$

The variance of $\Delta\theta_i$ in regions 1 and 3 can be calculated as
$$\text{Var}(\Delta\theta_i|A_1) = \text{Var}(\Delta\theta_i|A_3) = \text{Var}(2\beta) = 4\text{Var}(\sqrt{2x}) = 4\left(x_0 - \frac{8x_0}{9}\right) = \frac{4}{9}\frac{n_d}{n_m} \tag{S12}$$

The expectation of $\Delta\theta_i$ in region 1 and region 3, due to the symmetry, can be written as
$$E(\Delta\theta_i|A_1) = \frac{4\sqrt{2}}{3}\sqrt{\frac{n_d}{n_m}}, \quad E(\Delta\theta_i|A_3) = -\frac{4\sqrt{2}}{3}\sqrt{\frac{n_d}{n_m}} \tag{S13}$$

In region 2, the angle deflection is also $\Delta\theta_i = 2\beta$ (Figure S10), where the angle $\beta$ has the same form as Eq. (S1). The variance has a similar form to Eq. (S8).
$$\text{Var}(\Delta\theta_i|A_2) = \text{Var}(2\beta) = 4\text{Var}\left(\theta_i - \arcsin\left(\frac{n_m}{n_r}\sin(\theta_i)\right)\right) \tag{S14}$$

The expectation of the angle deflection in region 2 (Figure S10) is
$$E(\Delta\theta_i|A_2) = 0 \tag{S15}$$

due to the symmetric distribution of the rays. Therefore, based on Eq. (S4), we can get the total angle deflection variance through a single circular reinforcement as
$$\text{Var}(\Delta\theta_i) = \frac{2}{9}\frac{n_d^2}{n_m^2} + 4\left(1 - \frac{n_d}{n_m}\right)\text{Var}\left(\theta_i - \arcsin\left(\frac{n_m}{n_r}\sin(\theta_i)\right)\right) + \frac{2}{9}\frac{n_d^2}{n_m^2} + \frac{16}{9}\frac{n_d^2}{n_m^2} + 0 + \frac{16}{9}\frac{n_d^2}{n_m^2} \tag{S16}$$

It can be rewritten as
$$\text{Var}(\Delta\theta_i) = 4\left(1 - \frac{n_d}{n_m}\right)\text{Var}\left(\theta_i - \arcsin\left(\frac{n_m}{n_r}\sin(\theta_i)\right)\right) + 4\frac{n_d^2}{n_m^2} \tag{S17}$$

For a single circular reinforcement, the rays are uniformly distributed along $x$ axis. It means that $\sin(\theta_i)$ follows a uniform distribution $\sin(\theta_i) \sim U\left[-\frac{n_r}{n_m}, \frac{n_r}{n_m}\right]$. The calculation of $\text{Var}\left(\theta_i - \arcsin\left(\frac{n_m}{n_r}\sin(\theta_i)\right)\right)$ can also resort to a numerical way (Supporting Information S.VII).

## S.III. Variance decomposition

Given a random variable $\Delta\theta_i$, its variance is written as
$$\text{Var}(\Delta\theta_i) = E(\Delta\theta_i^2) - E(\Delta\theta_i)^2 \tag{S18}$$

Assume a partition of the outcome space into mutually exclusive and exhaustive events $A_k$ ($k = 1,2,3$). According to the law of total expectation, we have
$$E(\Delta\theta_i^2) = \sum_{k=1}^{3} P(A_k)E(\Delta\theta_i^2|A_k) \tag{S19}$$
$$E(\Delta\theta_i)^2 = \left(\sum_{k=1}^{3} P(A_k)E(\Delta\theta_i|A_k)\right)^2 \tag{S20}$$

Therefore, the variance of $\Delta\theta_i$ is derived as
$$\text{Var}(\Delta\theta_i) = \sum_{k=1}^{3} P(A_k)E(\Delta\theta_i^2|A_k) - \left(\sum_{k=1}^{3} P(A_k)E(\Delta\theta_i|A_k)\right)^2 \tag{S21}$$

According to the definition of conditional variance, we have
$$\text{Var}(\Delta\theta_i|A_k) = E(\Delta\theta_i^2|A_k) - E(\Delta\theta_i|A_k)^2 \tag{S22}$$

The variance $\text{Var}(\Delta\theta_i)$ is further decomposed as
$$\text{Var}(\Delta\theta_i) = \sum_{k=1}^{3} P(A_k)\text{Var}(\Delta\theta_i|A_k) + \sum_{k=1}^{3} P(A_k)E(\Delta\theta_i|A_k)^2 - \left(\sum_{k=1}^{3} P(A_k)E(\Delta\theta_i|A_k)\right)^2 \tag{S23}$$

## S.IV. Variance of $\sin(\Theta_{end})$

For simplicity, supposing that $\Theta_{end}$ follows a normal distribution with an expectation of $\mu$ and a standard deviation of $\sigma$, we have

$$\text{Var}(\sin(\Theta_{end})) = \text{E}(\sin^2(\Theta_{end})) - \text{E}(\sin(\Theta_{end}))^2 \tag{S24}$$

where

$$\text{E}(\sin^2(\Theta_{end})) = \frac{1}{2}(1 - \text{E}(\cos(2\Theta_{end}))) \tag{S25}$$

The key is to obtain $\text{E}(\cos(2\Theta_{end}))$ and $\text{E}(\sin(\Theta_{end}))$. To this aim, we derive the expectation of $e^{ia\Theta_{end}}$

$$\text{E}(e^{ia\Theta_{end}}) = \text{E}(\cos(a\Theta_{end}) + i\sin(a\Theta_{end})) = \text{E}(\cos(a\Theta_{end})) + i\text{E}(\sin(a\Theta_{end})) \tag{S26}$$

where $a$ is a constant. Based on the characteristic function, we have

$$\text{E}(e^{ia\Theta_{end}}) = e^{ia\mu - \frac{1}{2}a^2\sigma^2} = (\cos(a\mu) + i\sin(a\mu))e^{-\frac{1}{2}a^2\sigma^2} \tag{S27}$$

It yields to

$$\text{E}(\cos(ax)) = \cos(a\mu)\, e^{-\frac{1}{2}a^2\sigma^2} \tag{S28}$$

$$\text{E}(\sin(ax)) = \sin(a\mu)\, e^{-\frac{1}{2}a^2\sigma^2} \tag{S29}$$

Consequently, we can obtain that

$$\text{E}(\sin^2(\Theta_{end})) = \frac{1}{2}(1 - \text{E}(\cos(2\Theta_{end}))) = \frac{1}{2}\left(1 - \cos(2\mu)\, e^{-2\text{Var}(\Theta_{end})}\right) \tag{S30}$$

$$\text{E}(\sin(\Theta_{end})) = \sin(\mu)\, e^{-\frac{1}{2}\text{Var}(\Theta_{end})} \tag{S31}$$

Finally, the variance $\text{Var}(\sin(\Theta_{end}))$ is calculated as

$$\text{Var}(\sin(\Theta_{end})) = \frac{1}{2}\left(1 - \cos(2\mu)\, e^{-2\text{Var}(\Theta_{end})}\right) - \sin^2(\mu)\, e^{-\text{Var}(\Theta_{end})} \tag{S32}$$

When the rays are vertically incident on the sample surface, we have $\mu = 0$. The variation becomes to

$$\text{Var}(\sin(\Theta_{end})) = \frac{1}{2}\left(1 - e^{-2\text{Var}(\Theta_{end})}\right) \tag{S33}$$

## S.V. Expectation of the scattering

The location of the fibers is assumed to be random. The expectation of the angle change inside the transparent composite is

$$\text{E}(\Theta_{end}) = \sum_{j=1}^{\frac{M}{2}} \Delta\theta_{i,j} = \frac{M}{2}\text{E}(\Delta\theta_i) \tag{S34}$$

Due to the symmetry of each fiber, the expectation of the ray orientation change of each fiber $\text{E}(\Delta\theta_i)$ is 0. Therefore, we have

$$\text{E}(\Theta_{end}) = 0 \tag{S35}$$

Additionally, assuming the entry surface and the exit surface are parallel, the emergent ray and incident ray are expected to follow exactly the same orientation.

## S.VI. Approximation of the Fresnel law with Snell's law

The reflected and refracted ratios are the power reflection coefficient (or reflectivity) R and power transmission coefficient T=1-R. The power reflection coefficient R is calculated as

$$R = \frac{1}{2}(R_s + R_p) \tag{S36}$$

where $R_s$ is the reflectance for s-polarized light

$$R_s = \left|\frac{n_1 \cos\theta_i - n_2 \cos\theta_t}{n_1 \cos\theta_i + n_2 \cos\theta_t}\right|^2 \tag{S37}$$

$R_p$ is the reflectance for p-polarized light

$$R_p = \left|\frac{n_1 \cos\theta_t - n_2 \cos\theta_i}{n_1 \cos\theta_t + n_2 \cos\theta_i}\right|^2 \tag{S38}$$

where incident and refracted ray angle are $\theta_i$ and $\theta_t$, respectively. The refractive index of the two optical media are $n_1$ and $n_2$, respectively. It is reasonable to assume that the rays are uniformly distributed in space. As shown in Figure S8, the $x$ coordinate ($x$ axis is vertical to the incident rays) of the rays should be uniformly distributed. In the transparent composites, the RIs for the two compositions are very close. Assume $n_1$, $n_2$ is 1.5, 1.52 (or the converse), respectively. The reflectivity R versus $x$ coordinate is shown in Figure S11. The reflectivity is only obvious in a very limited range. It means that we can reasonably assume either only reflection or only refraction occurring at the interface.

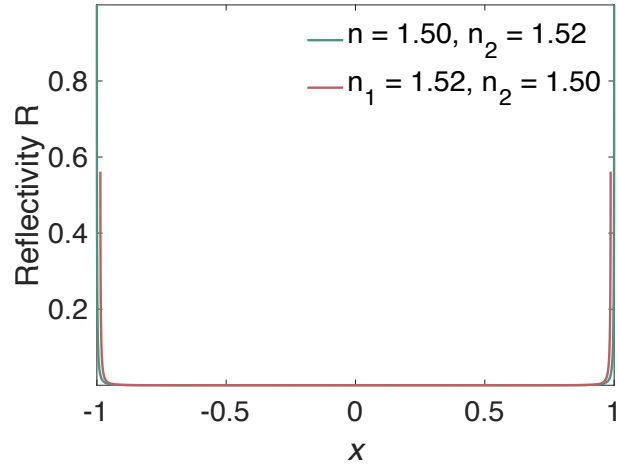

**Figure S11.** The reflectivity R for the incidental rays with different $x$ coordinate; it is almost 0 in most of the regions.

## S.VII. Numerical solution of the variance

There is no analytical solution for the variance of

$$\mathrm{Var}\left(\theta_i - \arcsin\left(\frac{n_m}{n_r}\sin(\theta_i)\right)\right) \tag{S39}$$

However, it can be obtained numerically. $\sin(\theta_i)$ follows uniform distribution in [-1, 1] when $n_m < n_r$ and in $\left[-\sin^{-1}\left(\frac{n_r}{n_m}\right), \sin^{-1}\left(\frac{n_r}{n_m}\right)\right]$ when $n_m > n_r$. Therefore, we can uniformly sample $\sin(\theta_i)$ in the domain with a constant step, such as 0.00001. For each value of $\sin(\theta_i)$, we can calculate the angle $\theta_i$. Finally, we can determine the variance $\mathrm{Var}\left(\theta_i - \arcsin\left(\frac{n_m}{n_r}\sin(\theta_i)\right)\right)$ numerically.

## S.VIII. Backward ray tracing for virtual image rendering

The backward ray tracing method was used to numerically render the image blurring effect through a GFRP or TW composite with a given thickness. For simplicity, we neglected the scattering along the fiber orientation. Suppose that the camera has a focal length of $f$. The sample was placed in front of the lens at an objective distance of $F_2$. The image distance is calculated by Eq. (S40) using the classical pinhole camera model (Figure S12).

$$\frac{1}{F_1} + \frac{1}{F_2} = \frac{1}{f} \tag{S40}$$

A transparent composite with a thickness of $T$ is put between the lens and the object. The gap between the object and the transparent composite is $d$. Suppose a pixel on the sensor at point A. Multiple rays, such as $\vec{AB}$ and $\vec{AC}$ can be inversely traced from that pixel to the lens at different locations at the point $B = (x_1, y_1, 0)$ and $C = (x_2, y_2, 0)$. The rays will be refracted at the lens plane and then focus at a given point $F$ on the object if there is no material between the lens and the object.

Due to the scattering of the transparent composite, the rays $\vec{AB}$ and $\vec{AC}$ will arrive at the point $I$ and $J$, instead of $F$ on the object. Suppose that the pixel $A$ has a coordinate of $(u, v, -F_1)$. Vector $\vec{AO}$ can be written as

$$\vec{AO} = (-u, -v, F_1) \tag{S41}$$

Then, we can calculate the coordinates of point F:

$$F = \frac{F_2}{F_1} \cdot \vec{AO} \tag{S42}$$

The coordinate of D is calculated as

$$D = B + \frac{F_2 - T - d}{F_2} \cdot \vec{BF} \tag{S43}$$

where $\vec{BF}$ can be easily determined since we know the coordinates of B and F. The key is to find the exit point and the emergent ray orientation on the opposite surface of the transparent composite. For simplicity, the ray tracing in the complex microstructure of transparent composites is avoided. Instead, we trace the ray in the composite by assuming a constant refractive index in the composite. We can neglect the scattering in the transparent composite to determine the location of G, since the sample thickness is assumed to be think. The coordinate deviation due to scattering is small in such a small thickness. Most of the image blurring is a result of the angle scattering. The coordinate of G on the light path BD can be easily obtained using Fresnel's law.

The angle scattering significantly affects the image blurring. Suppose the angle of the ray at $G$ has its projection along x axis and y axis (along fiber orientation) as $\theta_x$ and $\theta_y$, respectively. The expectation of the angle change after passing through the composite is 0. Therefore, the rays are expected to follow the original direction before and after passing through the composite. Their expectations are $\mu_x$ and $\mu_y$, which are the same as the projections of $\vec{BD}$ on x and y axes. We neglect the scattering along $y$ direction. The angle is assumed to follow a Gaussian distribution $\sin(\theta_x) \sim N(\sin(\mu_x), \sigma(\sin(\theta_e)))$. The standard deviation $\sigma(\sin(\theta_e))$ can be calculated using the analytical model we proposed in this work. Therefore, we can sample one angle $\theta_x$ from this distribution.

Based on the coordinate at G and the orientation of the emergent ray, we can obtain the coordinate at I on the object. In the same way, for all the rays coming from the pixel $A$, we can trace their intersection with the object. The average of all those intersection points on the sensor yields the image intensity at the pixel $A$.

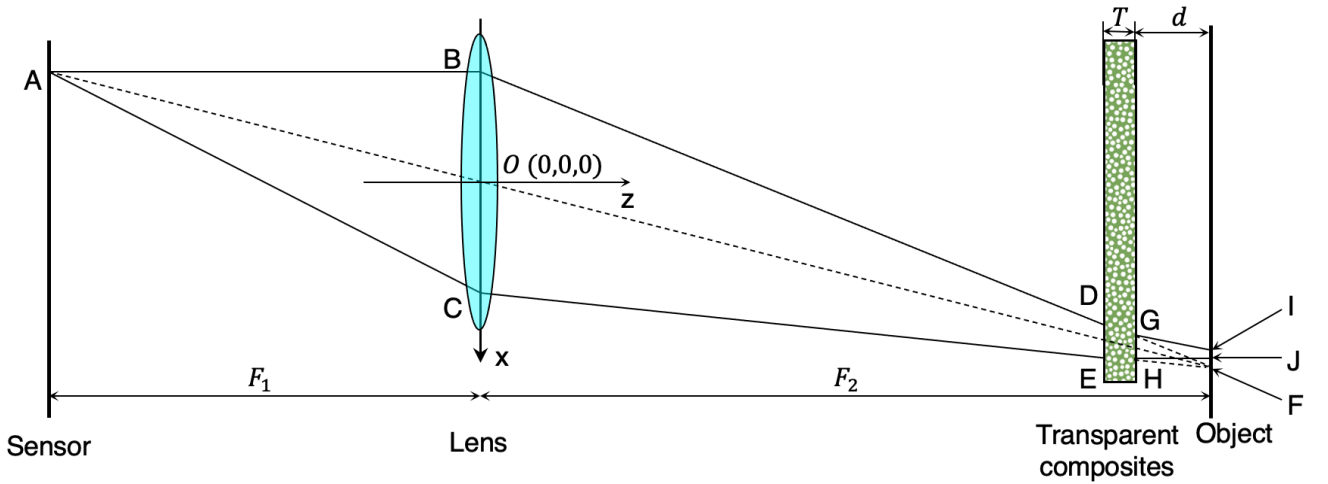

**Figure S12.** Principal diagram of backward ray tracing in TW.

## S.IX. Supplementary videos

**Video S1.** Haze versus EAIN = $M^k n_d$ fitted by a 10th-order polynomial function, where the green band is the residual band: (left) the relationship between haze and EAIN for different $k$, (right) the root mean square error in the fitting.